\documentstyle{article}

\begin{document}

\begin{center}
{\huge\bf Quantum Cohomology and all that}
\end{center}

\vspace{1cm}
\begin{center}
{\large\bf 
F.GHABOUSSI}\\
\end{center}

\begin{center}
\begin{minipage}{8cm}
Department of Physics, University of Konstanz\\
P.O. Box 5560, D 78434 Konstanz, Germany\\
E-mail: ghabousi@kaluza.physik.uni-konstanz.de
\end{minipage}
\end{center}

\vspace{1cm}

\begin{center}
{\large{\bf Abstract}}

We found a quantum cohomology, homology of quantum Hall effect which arises as the invariant property of the Chern-Simons theory of quantum Hall effect and showed that it should be equivalent to the quantum cohomology which arose as the invariant property of topological sigma models. This isomorphism should be related with an equivalence between the supersymmetric- and quantization structures in two dimensional models and, or with an equivalence between topological sigma models and the Chern-Simons theory by the methode of master equation.
\end{center}

\begin{center}
\begin{minipage}{12cm}

\end{minipage}
\end{center}

\newpage
{\large{\bf Introduction and summary}

\bigskip
Recently we showed the existence of a non-trivial "quantum" cohomology class 

$H^2_Q (\Sigma ; U(1)) \neq \emptyset$ on a $2-D$ manifold $\Sigma$ which is related with the quantum Hall effect (QHE) and which is absent in the classical case \cite{mein}. The importent point about this cohomology group is that its dual "quantum" homology $ H_2 (\Sigma ; R)$ is realized by experimental results of QHE in the sense that the $2-D$ quantum Hall-samples, i.e. $2-D$ electronic systems under quantum Hall conditions possess no "classical" (one dimensional) boundary \cite{meinedg}. 

According to this QHE-model of "quantum" cohomology the empirical boundarylessness of sample's manifold $\Sigma$ under QHE results in the geometrical fact that there are closed but not exact $two-forms$ defined on such manifolds, i. e.  $dF = 0, F\neq dA$. Thus, there are also non-trivial sheaf cohomologies $H^1 (\Sigma; A) \cong H^0_Q (\Sigma ; F) \neq \emptyset$. Furthermore, in view of Poincare duality and Hodge theorem one has $H^2_Q \cong H^0_Q \cong Harm_Q^0$ where the last one can be considered as the so called Floer cohomology \cite{Nash}. This is a cohomology related with our "quantum" cohomology, which is given by $\Delta_Q Harm^0 = E_0 Harm^0$ \cite{mein}, where $\Delta_Q$ is a deformation of the Laplacian $\Delta_Q := \Delta - E_0$ or $\Delta _Q := \Delta + O(\hbar)$ and $E_0$ is the ground state energy.

However, as the use of "quantum" indicates, this kind of quantum theoretically present but classically absent cohomology has to be proved to be equivalent to the quantum cohomology(QC)\cite{buch origin}. Recall also that it is already proved that QC which is a property of toplogical sigma models is equivalent to Floer cohomology \cite{sadov}, whereas the Floer cohomology as well as our QHE-cohomology is a property of a Chern-Simons theory without any explecit relation to the supersymmetry of topological sigma models \cite{mein}. Thus, also from the point of view of Floer cohomology the QC results from a quantized symplectic Chern-Simons structure $(\int A \wedge dA)$, i. e. from a quantum topological structure without supersymmetry of original models \cite{Witva}. Therefore, the proved equivalence between the QC and Floer cohomology supports our stand point that QC is not a result of supersymmetry, but it should be a result of the more general quantum structure of original models.

\bigskip
Moreover, in view of the fact that our QHE model is based on the Chern-Simons theory \cite{IQFQundmein} we will show that by the equivalence of Chern-Simons and topological sigma models (see below) our QHE-cohomology becomes equivalent to the QC of topological sigma model.
Althuogh, in view of the mentioned relation between the quantization structure and supersymmetry in $2-D$ models of QC \cite{mein}(see also below)  it is indeed difficult to distinguish between them in complicated models. Nevertheless, there are some principal questions with respect to the origin of QC, so that their clarification will be helpful for a physical understanding of QC and also for the mentined equivalence.

\bigskip
As another support for our stand point about QC let us mention that, as it was discussed in \cite{mein}, the main difference between classical and quantum phase spaces is the simply connectedness of the first and the multiply connectedness of the second, which induces different homological structures in these cases. In other words, the classical phase space has genus zero $g = 0$, whereas the quantum phase space as a multiply connected manifold has $g \neq 0$.
This property of multiply connectedness is related with the concept of quantization in the sense that quantization can be considered as a transformation from mutivalued functions into the single valued one (see the quantization of angular momentum ), which is also related with the concept of Novikov-ring ($\sim$ multivalued functions) \cite{mein}.

We  that the existence of the fundamental uncertainty relations in QM is equivalent to its $g \neq 0$ property \cite{mein}. This stand point can be supported with respect to the integrable systems in view of the fact that the classical limit $(\hbar \rightarrow 0)$ in these cases implies the restriction of genus of moduli space to zero $(g = 0)$ \cite{diieck}. Thus, in view of the equivalence between the mentioned moduli space and the reduced phase space ( see also below) the classical limit results in simply connected classical phase space of genus zero.

We mean here always by phase space the equivalent classes in phase space with respect to the relevant "canonical transformations". 

It is in view of such a deformation of homological/cohomological structure of phase space, as a result of multiply connectedness according to the quantization of phase space, that one may consider the QC as a deformation of the trivial ccohomology of classical phase space: one speaks from the division of phase space of a quantum system into cells with the area $(\propto \hbar)$. In view of quantization of system $(\hbar \neq 0)$, such cells and thereby loops surrrounding them can not be shrunk to a point \cite{mein}. In the same sense, the quantized or multiply connected phase space $(\hbar \sim g \neq 0)$ may have a classical or simply connected limit, i. e. the classical phase space $(\hbar \sim g = 0)$.

\bigskip
{\large{\bf Supersymmetry and Quantization}

\bigskip
Let us first mention that one main circumstance which seems to support the equivalence between supersymmetry and geometric quantization can be seen in Witten's approach to the supersymmetric generalization of Morse theory. Recall that Morse theory is related with the very general invariant geometric properties of manifolds under consideration, e. g. with their cohomologies, whereas supersymmetry is a special physical model which is not realized in the real physical world. Therefore, the only realistic conclusion can be: that supersymmetry is a special description of a more general structure, which we consider to be the geometric quantization \cite{mein}.

Thus, the main question here is the dependence of QC on supersymmetric structure of model, because the cited original models \cite{Witva} have this structure. 

As it was mentioned \cite{mein} this question has for us a negetive answere, because the fundamentality of QC as {\it the} quantum topological structure prevents any dependense of this kind. In other words, QC is of more fundamental and general nature, spacially in view of its invariant structure, so that it does not need to be bounded to the class of supersymmetric models.

To be precise, let us recall that in both original models parallel to the supersymmetry there is the Poincare duality of $two-forms$ and $zero-forms$ which is trivial in two dimensions. Therefore, as it is mentined in Ref. \cite{mein} the supersymmetry can be considered as a manifestation of quantization in such $2-D$ models. Recall that such supersymmetry, as a relation between quantum objects $bosons$ and $fermions$ in these models is described by a differential structure via an exact and closed $one-form$, i. e. $\psi \propto \delta \phi; \ \delta \psi = 0 $, which is equivalent to $H^1 = 0$ as it is used in the original models of QC \cite{Witva}. Abstractly, this is equivalent to the existence of a flat $U(1)$ connection on a bundle over a $2-D$ manifold which describes a quantization of the system under consideration according to the geometric quantization. Hence, in quantized case such flat connections have their invariant (global) contribution to the quantum phase of the system under consideration. Thus, these flat connection and the moduli space of mentioned original models are comparable with the flat connection of geometric quantization and with the phase space respectively (see also Ref. \cite{mein}), whereas the quantization relation of Bohr-Sommerfeld-Wilson \cite{wood} is a topological invariant which can be compared again with the basic topological invariants of supersymmetric models which is defined over the moduli space of flat connections \cite{Witva}. Moreover, the flat character of such a connection is related with the vanishing first Chern class $c_1 = 0, \ c_1 \in H^2$ which arises in Calabi-Yau manifolds, hence the $c_1 = 0$ property on a polarized symplectic manifiold is equivalent to the flatness of mentioned connection. 
The reason is simply that the covariant description of polarization manifests the existence of flat connections. Equivalently, in Calabi-Yau $3-folds$ which is an example of the extended phase space (see also below), one has according to the Poincare duality the flatness $H^1 \cong H^2 = 0$. 

In the quantized theory, the above discussed questions of flat connection are equivalent to the fact that here one has to do, instead of functions, with sections on a (complex) line bundle. Therefore, the above supersymmetric relation can be rewritten in its covariant (quantized) form as: 

$\psi \propto ( \delta - A ) \phi$ 

$F(A):= ( \delta - A ) \wedge ( \delta - A ) = 0$ \qquad,

or

$(\delta - A) \psi = 0$

where $A$ is the flat $U(1)$ connection which is related with the Calabi-Yau structure. 

In view of the flat character $\delta \phi = A \phi$, which stablishes a supersymmetric equivalence between $\psi \sim \delta \phi$ and $A \phi$.

It is importent also to mention here that in view of the self-duality of $one-forms$ in the $2-D$ case the flatness of the $U(1)$ connection, which manifests the quantization, is equivalent to the Lorentz gauge. Furthermore, as it is dicussed above the quantization becomes equivalent to the complex covariant description of mechanics, where wave functions are covariantly constant sections on the related (complex) line bundle.

\bigskip
Related with that question there are also questions about classical cohomology(CC) and QC, namely:

1) Is QC really the cohomology of the quantized phase space of the model under 

consideration ?

2) What is the cohomology on the classical phase space (CC) ?

3) what is the relation between the CC and QC ? 

We answere the first question positively and mention that the moduli space of flat connections ( $H^1 = 0$ ) in original models, where the relevant quantum invariants are calculated thereon \cite{Witva}, is indeed the polarized phase space of that systems.

The second question has to be answered with: trivial, if it concerns the true polarized classical phase space, i. e. without any quantum relicts. The reason should be related through the homology group with the simply connectedness of the true classical phase space, i. e. with its contractibility to a point.

For the third question we give a general answere which become more precised here, although it was introduced already in our previous work \cite{mein}. The relation in question is the quantization which is a "topological" property in the sense that the geometric quantization does not depend on the metric but it depends only on the symplectic structure or its complexification via almost comples structure on the phase space. Let us mention further that this relation could be described also by "deformation" in the sense that quantization is a deformation of the classical structure of mechanics. Moreover, in the same sense the relation in question 3) can be considered as the "quantum"  deformation of the first Chern class, i. e. from $H^2_{cl} = \emptyset \rightarrow H^2_Q \neq \emptyset$.

We mentioned already \cite{mein} that the crucial experimental fact about the absence of classical boundary of a $2-D$ system $\partial \Sigma \neq \emptyset$ and its relation with $H^2 (\Sigma; U(1)) \neq \emptyset$ is realized by the QHE situation where $\Sigma$ is the QHE-sample and the $H^2$-representant is the constant strong magnetic field strength $dF= 0$ which is applied on the QHE-sample. On the one hand, it is well known that the only consistent theoretical models of QHE are based on the $(2+1)-D$ Chern-Simons theory \cite{mein} which are topological field theories and have to be quantized in order to explain the QHE. Hence, the QHE is a quantum-topological effect. Thus, as a theory of QHE the quantized Chern Simons theory possess both ingredients of QC, namely the topological structure and the quantum structure which in our opinion are enough for QC.
Recall also our remark in Ref. \cite{mein} that experiments on edge currents and edge potential drops which are QHE-experiments support the fact about the absence of a classical one dimensional boundary of QHE-samples, but the presence of a quantal two dimensional boundary ring with a width related to their own magnetic length $l_B$. 

On the other hand, recall that only a quantized Chern-Simons theory can explain the $l_B$-ring as the boundary of QHE-samples related with the edge currents and quantum potential drops of flat U(1) potentials on the QHE-samples \cite{mein}, whereas the classical Chern-Simons potentials should be placed exactly on the classical one dimensional boundary of sample \cite{witten}. Thus, a natural origin of our quatum theoretically non-trivial but classically trivial cohomology is given by the Chern-Simons theory of QHE where the classical theory results in flat potentials defined on the classical boundary of sample, which determines such a boundary $(\sim H_{cl}^2 = \emptyset)$, whereas the quantum Chern-Simons theory results in potential drops indicating the absence of a one dimensional boundary for the samples $(\sim H_Q^2 \neq \emptyset)$.

\bigskip
{\large{\bf Topological Sigma Models and Chern-Simons Theory} 

\bigskip
As we conjectured \cite{mein} the phase space structure of QHE-system, i.e. of its Chern-Simons action functional is equivalent to the phase space structure of topological sigma-models. This is proved in a work using the master equation method of Batalin-Vilkovisky (BV) \cite{konts}. Here it is shown that the Chern-Simons theory in BV-formalism arises as a $2-D$ sigma-model with target space $\Pi {\cal G}$, where ${\cal G}$ stands for a Lie algebra and $\Pi$ denotes the parity inversion. Obviously in our case ${\cal G}$ is the $U(1)$ algebra ${\cal U} (1)$and we have to consider maps $\Sigma \rightarrow M = \Pi {\cal U} (1)$.

The equivalence between two type of topological models, i. e. Chern-Simons and sigma-model can be understood beter, if one recall the extended action functional for Chern-Simons theory \cite{AS k}.

Therefore, embeding our previous result $H^2_{(QHE)} \neq \emptyset$ \cite{mein} in its original theoretical background, i. e. in the Chern-Simons theory of QHE, we have $H^2_{(QHE)} \neq \emptyset$ as a result of quantized topological Chern-Simons/sigma-model which results in the standard QC. In other words according to the above discussed equivalence between the topological sigma- and Chern-Simons models, our QHE-cohomology is equivalent to QC.

\bigskip
To see the above mentioned equivalence via BV-approach in a more general way, i. e. between the $A$ and $B$ models one should recall the abstract classical mechanical foundation of that approach. We give here an equivalent approach which avoids the technicalities of the BV-approach:

Obviously, one is free to consider the phase space $X, \ {\{\pi (t), \kappa (t) }\}$ or the extended phase space $Y,\ {\{ \pi,\ \kappa, \ t}\}$ as the basic manifold of mechanics. The actual dimensions in both cases is the same $2n+1$, where we restrict ourselves to $n = 1$. However, in the first case the momentum and position variables $\pi$ and $\kappa$ are functions on the phase space and also of the time parameter $t$, whereas in the second case the last dependency is no more explicit. Accordingly, in the first case the $maximal \ cohomology$ depends on the symplectic $two-form \ (d\pi \wedge d \kappa)$, whereas in the second case it depends on a $three-form \ (d\pi \wedge d \kappa \wedge dt)$. This can be understood also in the following way that the $H^3 (Y; {\mathbf R})$ in the second case can be considered also as a $sheaf \ cohomology \ H^2 (Y, \Omega^1_0)$ which corresponds to the $H^2(X; R)$ of the first case, where the $\Omega^1_0$ stands for a $one-form \ (f dt)$. If we complexify $X$ and $Y$ manifolds in a suitable manner, then the $H^2 (X)$ goes over to the $H^{1,1}(X)$, whereas the $H^3(Y)$ can be written as the $H^{2, 1}(Y)$. Moreover, if one considers the $H^2 \cong H^1$ of $3-D$ manifolds, then the equivalence of that complex cohomologies $H^{2, 1}$ and $H^{1, 1}$ becomes more obvious. 

Furthermore, if we considere $X$ and $Y$ as Calabi-Yau manifolds, then the quantization of the first case which corresponds to the Heisenberg picture can be considered as equivalent to the $A-model$ of topological theories, whereas the quatization of the second case which corresponds to the Schroedinger picture can be considered as the dual $B-model$. For the later discussion recall also that in the second case $(\sim Y)$ the covariant version of the polarization $\partial_t \Psi = 0$ is given by $(\partial_t + \hat{H})\Psi = 0$, where the Hamilton operator $\hat{H}$ plays the role of a flat connection over the time manifold (see below and also \cite{mein}).

Nevertheless, we will consider again our previous statement according to which the QC should be a fundamental quantum property independent of special structure of models, where it arose originally \cite{Witva}.

To make this statement more and more plausible than by already introduced reasons let us recapitulate following facts:

a) The main properties of QC are its defining quantum and topological (invariant) properties.

b) It arose in models \cite{Witva} which have these properties, i. e. these models demonstrate QC in view of the fact that they manifest quantum topological (invariants). 

c) Thus, any model which manifests quantum topological (invariants) should demonstrate QC

d) Any kind of quantization should be equivalent to the canonical quantization which is based on geometric quantization which again is based on a toplogical invariant structure.

e) Therefore, the abstract (canonical) geometic quantization model should demonstrate also QC.

f) In other words only the geometric quantization structure of toplogical models \cite{Witva} should be responsible for the observed QC. 

\bigskip
We will demonstrate this line of arguments for an example in the appendix.

\bigskip

However, beyond the above discussed equivalence, we should compare also the structure of manifolds which underlies the $2-D$ TFT and the structure of a $2-D$ phase space which is complexified. The J-holomorphic curves which map the Riemann surface into the almost complex target space of TFT is the same as canonical transfomations in the phase space which maps the phase space into itself (automorphisms). Hence, the J-holomrphic curves are essentially the $2-D$ symplectic submanifolds of the target space. 
Of course one can consider also higher dimensional phase spaces which however are decomposed in view of quantization always into the $2-D$ quantum cells of the area $\hbar$. This seems to be related to the decomposition property of general many particle scatering amplitude into its $2-D$ subchannels \cite{reshetikin} which arises in integrable systems.

This complets the identification of the original TFT \cite{Witva} with the Chern-Simons theory or with the abstract canonically quantized phase space, where the main translation key is the flat one-form which is on the one hand the variationless spinor and on the other hand it is the flat $U(1)$ connection of the Chern-Simons theory or that of the geometric quantization.
 
\bigskip
{\large{\bf Mirror Symmetry and Polarization}

\bigskip
Another impertent hint about this type of generalization of QC which we try to construct comes from the essential relation between the QC and the mirror symmetry \cite{vaBuch}. Here, there are on the one hand the relation between the mirror symmetry and Fourier transformations in the phase space \cite{vaBuch} and on the other hand, we have the relation of mirror symmetry with the Abelian duality constructed via flat Abelian connections \cite{Morris}. Accordingly the mirror symmetry is related with the constrained mechanics and its Lagrange multipliers. The daul model is then a model for the Lagrange multiplier of the original model. 

As it is argumented below the dual and mirror structure is a structure of the polarized/constrained phase space of theory under consideration. We give herefore a fundamental example: Recall that the symplectic $one-form$ can be written instead of $ \theta = \pi d\kappa$ by $\theta = \displaystyle{\frac{1}{2}}(\pi d \kappa -  \kappa d \pi)$ because both models have the same symplectic $two-form \ (\omega := d \pi \wedge d \kappa)$ which has to vanish by polarization, hence its invariant integral can be considered as the "on-shell" action, or a part of "off-shell" action $\int \int \omega - dH \wedge dt$ of the general classical theory. 

Considering one of the two possible polarizations of phase space where the action becomes, beyond $t$, a function of $\kappa$ or $\pi$ only, then the other variable becomes the Lagrange multiplier of the original model which can be made to the fundamental degree of freedom, if one use the other polarization. These two possible models are dual to each other and the canonical transformation which relates these choises of polarization is the mirror symmetry. The quantization of the general model via integrality condition \cite{wood}
$\int\int\omega \propto \hbar$ which is performed by a flat $U(1)$ connection relates the mentioned duality to the Abelian duality of mirror symmetry \cite{Morris}. Recall further that $\pi$ and $\kappa$ are dimensionally and quantum theoretically inverse variables $\delta \pi \cdot \delta \kappa = \hbar, \ (\hbar = 1)$.

Therefore, the mirror structure of topological sigma models is the same as the polarization structure of the quantized phase space, whereas the so called "twisting"
of the first is related with the covariant polarization or with the covariant description of polarization of the second. 

On the other hand, the mirror symmetry manifests the equivalence between the topological sigma model and Ginzburg-Landau model, where the cohomology ring of one is mapped on complex ring of the others \cite{vaBuch}. Thus, genrerally one can formulate the same TFT by two field theories on two manifolds (of fields) which are dual of each others in the sense that the main parameter of one such manifold is dual or mirror or simply just the inverse parameter of the other manifold. However, sometimes where the polarization is not compatible with constraints one get a non-quantized classical theory. Recall also that the Ginzburg-Landau theory was concepted as a theory of superconductivity which is know considered as related with the QHE which again is described by Chern-Simons theory. Thus, if both QHE and superconductivity can be described by one and the same quantum theory, then the observed duality or mirror symmetry between these theories should be considered as a symmetry within one and the same quantum theory. Followingly, such a symmetry within a quantum theory can be only a symmetry of the quantized phase space which can be polarized in two mirror symmetric way, 
i.e. $\Psi(\pi , t)$ or $\Psi(\kappa , t)$.

\bigskip
Furthermore, it is obvious that topological properties are independent of local parameters of the manifold, thus one can find always locally different but topologically equivalent  manifolds. Of course these properties depend on topological properties of their manifolds, e. g. on cohomological or homological classes which are (in quantum theories) invariants describing their quantum numbers . The translation of this property to the case of quantum phase spaces or quantized manifolds which are line bundles over classical phase spaces or manifolds is in our opinion the essence of mirror symmetry, where the usual manifolds are changed to manifolds of fields which are again defined on usual manifolds. It is in this sense that the Floer cohomology seems to come from a quantization of supersymmetric sigma models which are based on maps between $3-D$ manifolds $M\rightarrow M$ instead of the usual sigma model maps $\Sigma \rightarrow M$ \cite{witsigma}. Among this local properties the main local ingredients of quantization is the polarization condition which is related with the parallelization of manifolds \cite{polparalel}, thus the quantization as a global or topologically invariant property has to be independent of polarization. In other words the quantum theory has to be symmetric with respect to various possible polarizations. This is what is called "mirror symmetry" in TFT/quantized CFT. Recall further that the Calabi-Yau structure $c_1 = 0$ is equivalent to the polarization of the same which is also equivalent to the constraint of the related classical Chern-Simon theory $(F = 0)$ (see below and also Ref. \cite{mein}). It is also in this sense that topological sigma-model on Calabi-Yau manifolds are equivalent to the quantized Chern-Simons theory. Since both are defined begining from a $2-D$ manifolds $\Sigma$ which is a Riemann surface, where in both cases one has to do with flat $one-forms$ which play also the role of flat $U(1)$ connection of quantization. Recall also that quantum theoretically the Chern-simons constraint $(F = 0)$ is a local condition in a region of the multiply connected phase space or moduli space of theory with $(F \neq 0)$ in its other region (see the Bohm-Aharonov effect), whereas the Calabi-Yau relation $(c_1 = 0)$ is a topological condition for such space.

\bigskip
In $2-D$ cases, where the fields $\phi$ are maps from a Riemann surface to the target space, there are to possible polarizations which are compatible with a Calabi-Yau structure of the target space. However, if one use a non-compatible polarization the resulting QFT should fail.

If one analyse the Witten's approach to mirror symmetry and duality \cite{witmir}, it becomes obvious that the different $A$ and $B$ models are based on different polarizations of the target space of the general theory $\partial_z \phi = 0$ or $\partial_{\bar{z}} \phi = 0$ which is related via supersymmetry with the polarization of the fermionic sections of the theory. So it results in view of the mentioned possibility of its covariantization by flat connections in the so called "twisted" bundles which have global effects in quantized theory. Of course in view of the supersymmetric structure of the original models the wohle structure of models is enriched in a complex manner. However, also in these cases one has the relation ($twisting \sim \ (covariant) polarization$). Therefore, the mirror symmetry which has to relate two different polarized models, i. e. two different polarized phase spaces with each other, has to be equivalent to the Fourier symmetry. Of course beyond the independency of quantization from the kind of polarization, the polarization has to be compatible with the present constraints of theory. The toplogical sigma-models are also in view of their equivalence with Chern-Simons theory constrained models. Therefore, their polarization should be compatible with their constraints oderwise the theory can not be quantized. Now the reason why $B$ theory does not make sense as a quantum theory is simply that it depends via complex structure of target space on the polarization, which is not allowed in a true quantization. It is the same, if the quantization of harmonic oscillator depends on the kind of complexification of its phase space. 

Recall again that the equivalent Chern-Simons theory is constrained by $F = 0$ which is the polarization of the complexified phase space and simultanously it is also the flatness condition of the $U(1)$ connection of quantization. 

On the other hand, the $A$ model does not depend on any complex structure, but on the cohomology class of the Kaehler form of target space or on the Kaehler class of the metric on target space. Taking the first version of dependency it does not matter for TFT/QFT in view of the globality of cohomology class. Also the second dependeny bares no danger for QFT/TFT because such a metric although it is a local property, however it is no essential property of the target space but only a trivial indication of the Hermiticity of an originally reel manifold which is now complexified. Therefore, the $A$ model in view of its independency of polarization albeit a trivial dependency is a good QFT.

In the case of classical $2-D$ manifolds the main invariant is the "area" related with the Euler characteristic, however in case of quantum manifolds we have to do with bundels over such classical manifolds, thus the invariants are defined not on the original usual manifold but on bundle space which are for examople of Chern or Kaehler type $(\int \int F \propto \hbar)$. 

In other words, the fundamental invariant of the quantized phase space is the area of its fundamental cell which is equal to $\hbar$ and to which all other invariants are proportional. In this sense quantum (field) theories defined on the momentum or position representations are equivalent by Fourier (mirror) transformations, where their momentum and position variables are inverse of each other in the $(\hbar = 1)$ normalization. Recall however that these representations are achived in quantum theory as well as in classical thaeory by polarization of the general phase space in two different ways. In this sense one can considere the two polarized/quantized phase spaces as mirror manifolds which results in one and the same topological, i. e. quantum theory. 

Moreover, on the one hand any polarization can be described in a covariant way introducing flat connections. On the second hand, in a constrained theory flat connections can be result from the constraints which is related with Lagrange multiplier methods. On the third hand, as it is mentioned above (line bundles) the geometric quantization is related with the existence of an Abelian flat connection which manifests the quantization by its loop integral which is proportinal to $\hbar$. Now, it is the interplay between these three properties of flat Abelian connections which results in the case of predestinated quantum theories of TFT or CFT type in various U(1) aspects of these theories which are related with their mirror aspects \cite{Morris},\cite {Witten u1}.

\bigskip

Comparing the quantized phase space of Chern-Simons theory with the abstract phase space ${\{\pi, \kappa}\}$ one observes a $SO(2)$ symmetry between the momentum- and position variables. Thus, the same is also possible in the usual phase space of one and the same theory, where one can use the canonical transformation $(\pi \rightarrow \kappa, \ \kappa \rightarrow - \pi)$ (see also above).

If we compare this symmetry and duality with the symmetry of strings on dual circles of circumference $L$ and $L^{-1}$, or if we compare it with theories where $\tau = \displaystyle{\frac{i R_2}{R_1}}$ is replaced by the area $\rho = i R_1 R_2$, then we find out that the main background of topological duality lies in the Poincare duality of $2-D$ manifolds to which the fundamental $2-D$ phase space also belongs. Here, according to $H^2 \cong H^0$ the topological structure of two forms is dual to that of zero forms. Followingly we have a duality between quantities which are components of two forms, i. e. of dimension $L^{-2}$ and dimensionless functions. The first are comparable with $\rho^{-1}$ and the last should be compared with $\tau$. Now the reason that this duality are considerable only in quantum topological theories and not in classical theories, lies in the impossibility of $(\delta l)^2 \rightarrow 0$ limit for the area in the (polarized) quantum (phase) spaces $(\int \int F \propto \hbar)$ which is related with the uncertainty principle or with non-vanishing ground state energy.

Another dualty which is considerable in QHE is within the phase space of the quantized Chern-Simons theory of QHE \cite{IQFQundmein}. Namely, one can register such a duality between the Hall conductivity and Hall resistivity $\sigma_H = (\rho_H)^{-1}$ on the one hand and on the other hand the duality between the space ${\{x_1, x_2}\}$ related with ${\{j_1, j_2}\}$ and the space of flat connections ${\{A_1, A_2}\}$, where each of them can be considered as the polarized phase space of the theory. Recall that the velocity Operator which is contained in $j_m$ contains $A_m$ which becomes quantized in the Chern-Simons theory of QHE. Recall further that one possible polarization in the phase space is a transformation from $\pi \rightarrow f(\kappa)$, after which the action functional and the wave function become a function of $\kappa$ only and the symplectic form vanishes as expected. Such a situation where all variables of the polarized phase space are functions of position variables ($x^m$ or $\kappa^m$) is given in the phase space of Chern-Simons theory ${\{A_1, A_2}\}$ or in the phase space of flux quantization $\int \int F \sim \int \int d\pi_m \wedge d\kappa^m \propto \hbar$, where $d\pi_m \sim \partial_n A_m (x^l) dx^n$ and $d\kappa^m \sim dx^m$.

It is in the above mentiond sense that dual structures results in one and the same quantum (topological) theory, where as an example one can look on the quantization of Hall-conductivity or of its reciprocal parameter the Hall-resistivity, where both results in one and the same topological QHE. Equivalently, one can consider as an example for this case the Flux quantization or the superconductivity as topological quantum theories.

\bigskip

Now what has this duality to do with our observation that in a quantum disc, i.e. a $2-D$ sample under QHE conditions, the impossibility of observation of the classical one dimensional boundary $\partial \Sigma$ results in the non-trivial cohomology $H_Q^2 \neq \emptyset$?

To see the relation first recall that the space of connections ${\{A_1, A_2}\}$ is related with the sample-disc as the configuration space ${\{x_1, x_2}\}$ via integrals of Ohm's equations or the mentioned Landau-gauge relations (see appendix). This configuration space is as good as the phase space of theory, therefore as a quantized space it is not possible to determine its boundary $(\partial \Sigma)_{Quantum} = \emptyset$. As we showed \cite{mein} this property results in a classically trivial but quantum theoretically non-trivial second cohomology. If one tries to determine a "quantum boundary" for this sample-disc, then one is obliged to determine it either by QHE experiments on edge currents or by the QHE experiments on edge potential drops. As it is shown \cite{mein} \cite{meinedg} the first type experiments will results in view of uncertainty relations in a boundary ring of the width $l_B$ and the second type experiments results in a boundary ring of the width $l_B^{-1}$. So the exixtence of such boundary rings show the impossibility of determination of the classical boundary under quantum conditions, whereas the duality of these widthes showes the imposibility of even a determination of a uniqe "quantum boundary" for the sample. It is also in this sense that the quantum boundary of our sample is undetermined or $(\partial \Sigma)_{Quantum} = \emptyset$. 

\bigskip
{\large{\bf Quantum Cohomology, Floer Cohomology and QHE-Cohomology}

\bigskip
We turn now to the relation between QC and Floer cohomology. As it is already clearified the main geometrical structure behind QC is that of symplectic geometry of the phase space which results in QC after quantization.
One importent hint in this direction which supports our point of view comes from the mentioned equivalence between the QC and Floer cohomology \cite{sadov}. Because Floer cohomology is on the one hand given on a model of quantized phase space ($\sim loop\ space \sim Hilbert \ space$ ) \cite{sadov} and on the other hand it is also related with a quantum deformation of the symplectic structure of its Chern-Simons action ($\sim Kaehler \ structure$) \cite{FloerWittsigmaunderklar}.
But it has no relation to an explicit supersymmetry, as like as our QHE-model, hence both are described by Chern-Simons theories. 

Nevertheless as it is mentioned in Ref. \cite{sadov} there are two obstacles to overcome to show that importent equivalence. We will discuss this overcoming from another point of view than in Ref. \cite{sadov}.

The first obstacle is that the Floer cohomology is naturally defined over the integer numbers ${\mathbf Z}$ and the QC is defined over comlex numbers and it depends on the Kaehler structure of the target space. This should be overcome, if one recalls that: 

1) The target space can be considered as the polarized phase space.

2) As it is argued in Ref. \cite{mein} the equivalence between the QC and Floer cohomology is of the type of equivalence between the Heisenberg's and the Schroedinger's representations of quantization. In other words the QC is a result of Kaehler quantization ($\sim Heisenberg \ picture$) which is concerned only with the pure phase space structure and its complexification. Whereas the Floer cohomology results from a quantization in Schroedinger picture where the phase of wave function results in integers $(\sim quantum \ numbers)$. Equivalently, one should consider the QC as the invariant structure on the manifold of quantum operators on the Hilbert space of theory which comes from a complexification of the symplectic (Hamiltonian) vector fields.
Whereas, the Floer cohomology should be considered as the invariant structure on the equivalent manifold of the state functions. We know that there is a correspondance between these two manifolds.

The second obstacle is that of holomorphic and pseudo holomorphic instantons  related with Hamiltonians $H = 0$ and $H \neq 0$ respectively which arise in QC and Floer cohomology respectively. In other words QC is independent of $H$ whereas Floer cohomology depends on $H \neq 0$. This one can be overcome , if we consider that as it was shown \cite{megeclas} the Hamiltonian $one-form$ $X_H dt$ is the flat connection of the fibre bundle formulation of symplectic mechanics over $t \in {\mathbf R}$ where classical mechanics becomes a gauge theory. Therefore, it can be gauged away within a consistent gauge transformation without destroying the gauge theory of classical mechanics \cite{megeclas}. The same procedure can be applied to quantum mechenics to achive its gauge theory, where the Hamilton operator becomes a flat connection $ (\partial_t + \hat{H} ) \Psi = 0$ (see above). Therefore, the $H = 0$ and $H \neq 0$ in classical or quantum mechanics are equivalent modulo a consistent gauge transformation. Followingly, all invariant structures which has to be gauge independent are independent of $H$ and in principle there is no need to proof this explicitely (see however the Floer's proof \cite{Floer}). One should consider this as a theorem: that symplectic invariants of a Hamiltonian system, e. g. its cohomologies, are always independent of the Hamitonian. 

Thus, turning to the second obstacle, two structures which results from these two equivalent situations $ H = 0$ and $ H \neq 0$ are also equivalent. Recall that a consistent gauge or a consistent gauge transformation in quantum mechanics means: $ (H \rightarrow H^{\prime}) \Leftrightarrow (\Psi \rightarrow \Psi^{\prime})$. Again this circumstance recalls the situation between the equivalent Heisenberg's and Schroedinger's picture, where in the first case the operators and states $\Psi$ are functions and constants of time respectively, whereas in the second case they are constants and functions of time respectively. So that indeed QC and Floer cohomology are the same invariant aspects described in two equivalent pictures of quantization , i. e. in the Heisenberg and in the Schroediger picures of quantum theory respectively.  

We showed further that the equivalence between our $H_Q^2 \neq 0$ and Floer cohomology of ground state should be given via Hodge and de Rham theorems on manifold of our interest, i. e. by $H^r \cong Harm^r$ and $H^m \cong H^0$. They results in our case in $H^2 \cong Harm^0$, where $Harm^0$ represents the ground state \cite{mein}. It is the same state which is BRST invariant according to its harmonic property.
Accordingly, also the of-shell and on-shell situations and the action of BRST operators discussed in \cite{sadov} can be translated into the fundamental quantum mechanical situations which are present in all quantized theories. One leson from this part is that also the internal degrees of freedom which are present in the Floer theory of non-Abelian Chern-Simons theory are not essential for the Floer cohomology or QC, but they enrich the cohomological ring structure via the rich structure of their moduli space.

\bigskip
{\large{\bf Fussion Ring and Quantum Cohomology Ring}

\bigskip
The last point to be clarified with respect to an identification between QC and QHE-cohomology is the ring structure of QC, i. e. $H^*$. It is given in the $2-D$ case given by $H^* = H^2 \oplus H^1 \oplus H^0$ where one uses in the original models $H^1 = 0$. Thus, one obtains $H^* = H^2 \oplus H^0$ which becomes according to the Poincare duality $H^* \cong H^2 \cong H^0$. As it is already mentioned this conditions becomes in our terminology the same as the flatness of quantization-connection. Here the more interesting relation can be constructed via fusion ring \cite{gepner} of quantum states.

The fusion ring structure for primary chiral fields in original models arises by their operator product $\phi_i \phi_j = N^k_{ij} \phi_k$. Thus, it is the ring structure of $zero-forms \ \phi \in \Omega^0$ which is given according to the Hodge theoprem by $zero-harmonics \ (Harm^0)$. Hence, one can define, in view of $(H^2 \cong H^0 \cong Harm^0 \cong Harm^2)$ an isomorphic ring structure for $two-forms \ \Omega^2$. Thus, the fusion ring of primary fields, i. e. $Harm^0$ should be the same as the ring structure of the Hodge dual $two-forms \ F \in \Omega^2$ or of $Harm^2$. Therefore, the cohomology ring on a $2-D$ manifold can be given by $H^*(\Sigma) = H^2\cong Harm^2$ as in our case or by its dual $H^*(\Sigma) = H^0 \cong Harm^0$ as in the topological sigma models \cite{Witva} depending on which field theory we define on $\Sigma$. In this sense our QHE-cohomology which has should be equivalent to that of the original models.

To see further relation to the fusion ring approach let us mention that one can identify the main ingredients of Jacobian ring approach \cite{gepner} with that of our QHE approach in the following way:

There are similar structures $P[x_i] \sim dF = 0$ and $(p_i) = \displaystyle{\frac{\partial V(x_i)}{\partial x_i}} \sim dA$, so that the Jacobian ring $(P[x_i]/p_i) = (P[x_i]/ \displaystyle{\frac{\partial V(x_i)}{\partial x_i}}) \sim (dF = o/F = dA)$ which is our non-trivial quantum $H_Q^2$ ring structure. Recall further that accordingly, the condition $\displaystyle{\frac{\partial V(x_i)}{\partial x_i}} = 0$ in Ref. \cite {gepner} becomes equivalent to the already mentioned flatness condition of the $U(1)$ connection.

Let us mention at the end that the central role played by Gromov-invariants in QC (see the Book in Ref. \cite {buch origin}), which results from a general (holomorphic) symplectic approach without explicite relation to supersymmetry supports our general appeoach to QC.

\bigskip
{\large{\bf Conclusion}

We know that it must be a relation between QC and quantum groups \cite{vafagepnermein}. To obtain it one may consider the following deformation

\bigskip
$[ T_+ \ , \ T_- ]_q : = \displaystyle {\frac{e^{\frac{\hbar T_3}{S}} - e^{-\frac{\hbar T_3}{S}}}{\frac{2\hbar}{S}}}$

\bigskip
In infinite volume limite $\displaystyle{\frac{s}{\hbar}} \rightarrow \infty$ this algebra goes over to the usual algebra of $T_i$ operators.

It is the same limit where the QC goes over to CC, i. e. to its trivializazation. In other words, the quantum group appears in the quantum limit of the classical algebra, i. e. in the limit $S = O(\hbar)$. One possiblity where this limit is given is just the $S \rightarrow \delta S$ limit for unitary operators of quantum theory $(\hbar = 1)$ . 

In other words, whereas the state functions $\phi_i = \displaystyle {e^{\oint A}} \phi_0 = \displaystyle {e^{\Theta_i}} \phi_0$ with $A:= A_m dx^m$ could result in commutative fussion ring structure of QC as mentioned above; The $\displaystyle ({e^{A_1 dx^1}})$ and $\displaystyle ({e^{A_2 dx^2}})$
operators corresponding to the infinitesimal transformation in the Hilbert space of $\phi_i$ have a non-commuting mutiplication for quantized $A_m$ according to $[A_1 \ , \ A_2] \propto \hbar$. This multiplication has the same quantum plane structure as $\omega_n \omega_{n+1} = q^2 \omega_{n+1} \omega_n$ \cite{faddeef} and it relates the QC of fussion ring type with the quantum groups on quantum plane. 

Therefore, considering the state functions as depending or representing pathes in quantized phase space, the state functions ("operators") depending on infinitesimal pathes obey quantum plane type relations as given above, whereas the state functions depending on loops in phase space obey commutative algebras of the mentioned fussion type.

We discuss these aspects and also the relation of our stand point to the Gromov's approach later.

\bigskip
Acknowledgment: I am thankful to Prof. Y. I. Manin and Prof. V. Puppe for stimulating discussions.

\bigskip
{\large{\bf Appendix}}

\bigskip
The Abelian Chern-Simons theory (model) can be considered as a topological sigma-model ( see above) which is comparable with the original models \cite{Witva}. The action of theory is given by 

\begin{equation}
k \int_{\Sigma \times R} A \wedge dA ,
\end{equation}
\label{one}

with $A$ representing $2-D$ flat U(1) connections. As we have shown \cite{mphysSC} this action can be written via Ohm's equations or their integral which is the Landau-gauge $A = B \wedge x$ for constant $B$ with $(dF = 0)$, in a simple form

\begin{equation}
\int A = \oint A 
\end{equation}
\label{two}

where we used $k^{-1} = B.S$ for a constant surface which should be represented by the surface of QHE sample.

Considering this topological invariant of $A$ as the same which is used for the geometric quantization via the flat $U(1)$ connection $A$, we arive in the very general model of the geometric quantization. Thus, quantized Chern-Simons or sigma-model theory is equivalent to the abstract theory of canonical quantization, i. e. to the geometric quantization. Recall also that by quantization the $A$ should be considered as a quantized operator. 

Recall also that the quantization of a non-abelian $SU(n)$ Chern-Simons theory shows that the quantizable- or quantized structure of the theory contains $(n^2 - 1)$ copies of the $U(1)$ Chern-Simons theory, although the two moduli spaces have different structures. Thus, with respect to the quantization a non-Abelian
Chern-Simons theory behaves like a collection of copies of Abelian Chern-Simons theory. Therefore, the QC of non-abelian cases should depends also only on their quantum structure which are copies of quantization of the Abelian case.

\bigskip
Footnotes and references

\end{document}